\documentstyle[twocolumn,aps]{revtex}

\begin{document}

\title{Multiplicative semiclassical dynamics and the quantization time}
\author{L. Kaplan \thanks{kaplan@physics.harvard.edu}
\\Department of Physics and Society of Fellows,\\ Harvard
University, Cambridge, Massachusetts 02138}
\maketitle

\begin{abstract}
We study smooth, caustic-free, chaotic
semiclassical dynamics on two-dimensional
phase space and find that the dynamics can be approached by an iterative
procedure which constructs an approximation to the exact long-time
semiclassical propagator. Semiclassical propagation all the way to the
Heisenberg time, where individual eigenstates are resolved, can
be computed in polynomial time, obviating the need to sum over an
exponentially large number of classical paths. At long times, the dynamics
becomes quantum-like, given by a matrix of the same dimension as the
quantum propagator. This matrix, however, differs both from the quantum
and the one-step semiclassical propagators, allowing for the study of
the breakdown of the semiclassical approximation.
The results shed light on
the accuracy of the Gutzwiller
trace formula in two dimensions, and on the source of long-time periodic orbit
correlations.
\end{abstract}

\vskip 0.2in

\section{Introduction}

Semiclassical methods have a long history dating back to the very
beginnings of quantum mechanics, and have provided insight into many
properties of quantum mechanical systems. These methods provide a bridge,
expressing quantum behavior
in terms of classical paths and their corresponding actions. In
integrable systems, the connection between quantum and classical
behavior is well understood through EBK quantization techniques, which
lead to an intuitive understanding of the quantum properties of these systems.
For a nonintegrable system, it is not nearly as clear how much of the
quantum behavior (e.g. spectrum, eigenstates, long-time dynamics,
transport) can be understood via semiclassical methods. One would
like to be able to separate out those features of the quantum behavior
which can thus be explained in terms of interference between classical paths
from the ``hard quantum effects", such as diffraction and tunneling.

Although the Van-Vleck formula describing short-time
quantum behavior in terms of classical paths has been around for many
years, it is only starting with the work of Gutzwiller\cite{gutz} that
it has been possible to try to understand long-time quantum properties
in terms of classical behavior. Much insight is given by the Gutzwiller
trace formula in the energy domain,
which relates the quantum spectrum to a sum over
classical periodic orbits.
However, the formula is a formal expression, that needs
to be resummed to get convergence in the limit where more and more orbits
are included\cite{berrykeat}. Furthermore, in practice summing over long
orbits (ones with period comparable to $\hbar$ times the
density of states of the system,
so that individual eigenstates can be resolved)
is not feasible, because the number of classical orbits grows
exponentially with time in a chaotic system.\footnote{Special cases are known
for which the long-time semiclassical dynamics can be computed exactly,
e.g. the cat map and geodesic flows on surfaces of constant negative
curvature.
However, the semiclassical evolution for these systems is in fact the same
as the quantum evolution because of their linearity
(or homogeneity in the case of constant negative curvature),
making them less interesting as a testing ground
for the general applicability of semiclassical methods.}
Cycle expansion methods, which
use the symbolic dynamics of the underlying classical system to express
long periodic orbits in terms of shorter ones have been very important
in this regard\cite{cycleexp}.

However, this still leaves open the question of the properties of
the semiclassical dynamics in the time domain. Much important work
here has been done  in systems like the stadium billiard and the baker's
map\cite{helleretc}. Numerical evidence was produced,
and theoretical arguments given, showing clearly that
the semiclassical approximation works well past the mixing time of the system,
where multiple stationary paths contribute to the quantum propagator, and
where the purely classical approximation (without interference effects)
breaks down completely. This was initially somewhat surprising, because
after the mixing time the classical dynamics begins generating structures
in phase space on scales smaller than Planck's constant, while the quantum
dynamics washes out any information on these scales. Thus the semiclassical
approximation tries to follow the quantum propagator by summing over an
exponentially large number of paths. This exponential proliferation of
classical paths prevents one from performing meaningful long-time 
quantum-semiclassical comparisons for small values of $\hbar$ (large
Heisenberg time). The fact that the quantum propagator effectively smears
out all of the sub-$\hbar$ structure
in phase pace, and thus contains an amount of information
which scales only as a power law in $\hbar$ suggests that some such
reduction may also be possible in the semiclassical calculation. This
is important because a priori it is not at all obvious (1) that the long-time
semiclassical dynamics converges in the sense of having well defined 
stationary states at the Heisenberg time,
or for that matter (2) that in those cases where
the semiclassical dynamics does converge, what it converges to actually
approximates the quantum dynamics, stationary states, etc. The breakdown
of semiclassical validity, in particular in the presence of diffraction,
discontinuities, and caustics, is also a somewhat controversial issue
which has been difficult to address in practice because of the
computational obstacles.

A key reason for the finiteness of long-time information in the quantum
case is the multiplicativity of the quantum propagator. In other words,
the long-time dynamics can be obtained by simply iterating the short-time
propagation (matrix multiplication in the case of a finite-dimensional 
Hilbert space). The semiclassical dynamics does not share this property
because concatenations of classical paths in general produce paths
that are {\it not} classical. Stationary phase integrals must be performed 
to obtain the long-time semiclassical propagator from shorter steps,
and the number of stationary phase points grows exponentially with time.
An interesting way to make the semiclassical dynamics multiplicative
by extending the space on which the multiplicative operator acts has
been described by Cvitanovi\'c and Vattay\cite{cvitvattay}.
It has also been seen that in the
special case of the baker's map, the long-time dynamics can be computed
with good accuracy in polynomial time using the Heisenberg uncertainty
principle and the  exponential decay of time correlations in
chaotic systems\cite{longbaker,iterbaker}. These methods (of which three are known
to the author) effectively collect together all contributions from
classical paths which come together on scales much smaller than $\hbar$.
Semiclassical amplitude thus collected can then be
propagated again, thus making the
resulting dynamics multiplicative. All the methods, however, made use of the
simple symbolic dynamics of the baker's map and its very special
structure in position and momentum space. Some of the ideas of consolidation
on sub-$\hbar$ scales,
however, may yet turn out to be fruitful in analyzing more
generic systems, with no such special structure\cite{iterbaker}.

In this paper, we take a somewhat different approach to the problem,
inspired by the fact that in the baker's map it was found\cite{iterbaker} 
that a good approximation to the long-time semiclassical
dynamics could be obtained
even if one periodically projects the higher-dimensional semiclassical
vector onto the $N$-dimensional quantum space. In other words, the true
semiclassical dynamics is evaluated exactly for $T_Q$
steps.\footnote{$T_Q$ is called the quantization time,
and can be thought of the
time domain analogue of the $T^*$ parameter in the 
spectral theory of Bogomolny and Keating\cite{bogokeat}. One should note,
however, that in periodic orbit theory, the critical time $T^*$ should
scale as the mixing time, logarithmically with $\hbar$, whereas our scale
$T_Q$ is of order one in units of the shortest periodic orbit. See the
discussion in Section\ref{conclusion}.}
Matrix elements are then taken
between quantum states, producing a matrix of the same dimension
as the quantum propagator. This matrix is then iterated to produce
an approximation to the long-time semiclassical behavior. Now for
$T_Q=1$ this is very reminiscent of Bogomolny's approach to
quantization\cite{bogoquan}. It should be noted, however, that for $T_Q=1$,
this procedure does {\it not} produce anything at all resembling the long-time 
semiclassical dynamics. In fact for small $T_Q$ one gets closer and closer
to the exact {\it quantum} dynamics, as can be seen by taking the limit $T_Q
\to 0$ and recovering the Feynman form of the quantum propagator. This is
good if one just wants to know the quantum answer, but not if one is
interested in understanding how quantum dynamics is affected by classical
properties, in the sense, for example, of the Gutzwiller trace formula.
What
is interesting here is that for $T_Q \gg 1$, but still much smaller
than the Heisenberg time, one can obtain a good approximation
to the actual long-time 
{\it semiclassical} dynamics, and thus study the validity (and
breakdown) of the semiclassical propagation. Detailed comparisons
become easily possible between quantum and semiclassical spectra, eigenstates,
long-time transport, and other properties.

Here we apply these ideas to a somewhat different class of systems, 
ones that lack the somewhat unpleasant ``cut and paste" discontinuities
of the baker's map, that are not piecewise linear,
and which are also more generic in the sense
of constituting a large family and allowing for continuous perturbations. These
systems, namely smooth automorphisms of the torus (e.g. kicked maps,
perturbed cat maps) lack a special symbolic dynamics and have no preferred 
basis, making them a good test case for the ideas outlined above.\footnote{
An important constraint is that we shall be looking exclusively here
at caustic free,
purely chaotic systems. Caustics, discontinuities, and mixed phase space are
all important subjects of inquiry and will be looked at in a future paper.}
It should be pointed out that the procedure presented in what follows does
not make use of detailed knowledge about the system at hand. In this way, it
may be more robust than, for example, the cycle expansion methods
in periodic orbit theory. Because no explicit use is made of periodic
orbits, it need not be the case here that long periodic orbits can be 
expressed in terms of short periodic orbits, for example. All that is needed
is that long paths can be constructed out of short paths, something that
is true in all but the most pathological cases.

The rest of this paper is organized as follows: we first briefly review the
classical, quantum, and semiclassical dynamics of the systems under
consideration. The iteration of semiclassical propagators is discussed,
and error
estimates are given as a function of propagation time, quantization time, and
$\hbar$, with particular emphasis on Heisenberg-time propagation.
Numerical tests follow, showing the convergence of the iterative approximation
as well as measuring the deviation from exact quantum and one-step
iterative results. Then the ``effective one-step semiclassical propagator"
is discussed, which with exponential accuracy describes long-time semiclassical
propagation, and which differs from the quantum and one-step semiclassical
matrices. Again, numerical findings are presented. Finally, the conclusion
treats some general questions and
addresses possible applications and extensions of the results obtained.

In a companion paper to this article~\cite{iterloc}, the methods are extended
and applied to the study of semiclassical dynamical localization in
classically diffusive systems, showing that in fact interference between
classical paths is sufficient to understand the end of quantum diffusion
in these systems.

\section{Theoretical analysis}

\subsection{Hard chaotic dynamics on a torus}

We will consider smooth, chaotic classical dynamics on a compact phase space,
satisfying the
Anosov property,\footnote{This means that at each phase space point the tangent
space can be decomposed into an exponentially expanding linear
subspace and an exponentially contracting linear subspace. For two-dimensional
phase spaces, there is one expanding and one contracting direction at each
point.}
and free from caustics in the coordinate system
of interest. To be specific, let us take the following family of discrete
area-preserving maps on a torus:
\begin{eqnarray}
\label{clasdyn}
p & \to & \tilde p =  p + mq - V'(q) \; {\rm mod} \;1 \\
q & \to & \tilde q = q + n\tilde p + T'(\tilde p) \; {\rm mod} \; 1\,.
\nonumber
\end{eqnarray}
The above dynamics can be obtained from the stroboscopic
discretization of a kicked
system\cite{kickmap} with a kick potential $-{1 \over 2}mq^2+V(q)$ applied once every 
time step
and a free evolution governed by the kinetic term ${1 \over 2}np^2+T(p)$.
Here $m$, $n$ are arbitrary
integers, while $V$, $T$ are smooth periodic functions. The system can also
be thought of as a continuous perturbation of the linear system (cat map) \cite{pertcat}
\begin{eqnarray}
\label{catmap}
p & \to & \tilde p = p + mq \; {\rm mod} \; 1 \\
q & \to & \tilde q = np + (mn+1)q \; {\rm mod} \; 1 \,. \nonumber
\end{eqnarray}

The Jacobian of the transformation in Eq.~\ref{clasdyn} is given by
\begin{equation}
\label{jacob}
J= \left[ \begin{array}{cc} 1  & m-V''(q)  \\ n+T''(\tilde p) \; &
    1 + (n+T''(\tilde p))(m-V''(q)) \end{array} \right] \,.
\end{equation}
We notice that for given integers $m$, $n$ we can choose the functions $V$
and $T$ such that the quantity $m-V''(q)$ is strictly greater than $0$
for all $q$ and similarly for $n+T''(\tilde p)$. (In words, we ensure
that the system everywhere looks locally like an inverted harmonic
oscillator.) This implies strict positivity
of all four entries in the Jacobian matrix, a property that is of course
preserved under iteration of the dynamics. Furthermore, positivity of both
off-diagonal entries implies hyperbolicity, because it ensures
$\det J > 2$. All such
systems
therefore provide examples of hard chaos, being free of integrable regions.
To see that they are free of caustics in either position or momentum space,
it is sufficient to note that the two off-diagonal entries of the Jacobian,
$\partial p(t) / \partial q(0)$ and $\partial q(t) /  \partial p(0)$ always
remain non-zero.

The quantization of kicked systems is straightforward and well-covered
in the literature\cite{kickmap}. We take $\hbar$ so that $N=1/2\pi\hbar$ has an
integer value. Then an $N$-dimensional position basis for the Hilbert
space is given by $|q_i\rangle$, where
$q_i=(i+\epsilon_0)/N$, $i=0 \ldots N-1$.
Similarly, the momentum
space basis is given by  $|p_j\rangle$, with allowed values
$p_j=(j+\epsilon_1)/N$, $j=0 \ldots N-1$.
$\epsilon_{0,1}$
form a family of possible quantization conditions (they correspond to phases
associated with circling the torus in the $p$ and $q$ directions,
respectively). The two bases are related by a discrete fourier transform.
The dynamics is now defined by the unitary $N \times N$ matrix
\begin{eqnarray}
U & = &
\exp{\left[i \left({1 \over 2} n\hat p^2 +
T(\hat p)\right)/\hbar\right]}
\nonumber \\ & \cdot &
\exp{\left[-i \left({1 \over 2} m\hat q^2 -V(\hat q)\right)/\hbar\right]}
\,,
\end{eqnarray}
where each factor is evaluated in the appropriate basis, and an implicit
forward and backward fourier transform has been performed.\footnote{As
in the linear (cat map) case, continuity of the potential and kinetic
terms (mod $\hbar$) requires that $N$ be chosen even, unless both $m$
and $n$ are even.} The key point is that $U$ here is just a matrix, and 
the long-time quantum evolution of the system is given by matrix
multiplication. Alternatively, the matrix $U$ can be diagonalized to find
the stationary properties (eigenstates and eigenvalues) of the system. All
this can be performed in polynomial time, and the total amount of information
contained in the quantum system is of order $N^2$. In particular, all of
the information is present in the one-step quantum propagator.

\subsection{Semiclassical dynamics}

We now consider the long time semiclassical dynamics as given by the
Gutzwiller--Van-Vleck
propagator\cite{gutz} evaluated between quantum mechanically allowed states.
In position space, the propagator has the form
\begin{eqnarray}
\label{gvv}
G_{\rm sc}(q,q',t) & = & \left[{1 \over 2\pi i\hbar}\right]^{d/2}
\sum_j \left|\det {\partial^2 S_j(q,q',t) \over \partial q \partial q'}
\right|^{1/2} \nonumber \\ & \times &
\exp\left[{i S_j(q,q',t) \over \hbar} - {i\pi \nu_j \over 2}
\right] \,,
\end{eqnarray}
where $S_j$ is the action for classical path $j$, the determinant corresponds
to the classical probability density
for going between $q$ and $q'$ via this path,
and the phase is given by the action,
corrected by the count of conjugate points $\nu_j$.
For any
given time $t$, this produces an $N \times N$ matrix $A_t$. The semiclassical
evolution must of course be evaluated in some coordinate system, such as
position or momentum, but once the matrix has been constructed it can be
rotated into any quantum basis that one finds convenient. Being a matrix
connecting quantum in and out states, $A_t$ looks like a quantum object,
but it is not unitary $(A_t A_{-t} \ne I)$
and does not satisfy  multiplicativity $(A_{t+t'} \ne A_t A_{t'})$. Thus
it is not a priori obvious, for example, that the semiclassical evolution
at long times converges to a well-defined set of eigenstates and
eigenvalues, i.e. $A_t\psi_n \approx e_n^t\psi_n$, or whether such eigenvalues
and eigenstates $e_n$, $\psi_n$
have any connection with those of the quantum dynamics.

Let us study the deviation from multiplicativity of the semiclassical dynamics
in a smooth chaotic system. We note first that for semiclassical propagators
in the time domain, 
\begin{equation}
\label{a2t}
A_{2t}(q'',q)=\int_{\rm sp} dq' A_t(q'',q')A_t(q',q) \,,
\end{equation}
where $\int_{\rm sp}$ indicates that the equality holds only if the
intermediate integration is performed by stationary phase\cite{gutzbook}.
If the integration is performed exactly instead of in the stationary phase
approximation, we obtain a relative error of order $\hbar$ in the answer.\footnote{
Of course, for large $t$, many stationary paths are
summed over on the right hand side
of Eq.~\ref{a2t}. The relative error in ignoring subleading terms in the
stationary phase expansion for {\it each}
such path is $O(\hbar)$. So the fractional error
in the full answer is also $O(\hbar)$, provided that the
errors add no more coherently
than the leading terms themselves.} This
error comes from higher-order terms in the stationary phase expansion. (In the
presence of caustics, the coefficient of the $O(\hbar)$ term can blow up
in certain regions of space, eventually dominating the semiclassical
evolution. Possibilities for handling this problem include uniformization
or choosing a different basis (e.g. a gaussian basis) for performing the
semiclassical calculation. To avoid these serious difficulties we will deal
throughout this paper with systems that are caustic-free in the chosen basis.)

For systems on a compact classical phase space, position and momentum values
are of course labeled by discrete integers, so strictly speaking the notion of
stationary phase integration is not well defined. To make sense of the
semiclassical dynamics, one must rewrite the sums
over topological classes (winding numbers) using the Poisson summation
formula, and evaluate the resulting integrals by stationary
phase\cite{pertcat}. This produces a semiclassical propagator defined only
on a discrete position (or momentum) grid.
However, in the limit of
small $\hbar$ (large $N$), the spacing between quantum basis
states vanishes, and
the discreteness of the quantum basis ceases to be physically significant.
This is true as long as all quantum structures in phase space are ``generic",
scaling as $\sqrt\hbar$ in both the $q$ and $p$ directions, while the spacings
$\Delta q$ and $\Delta p$ scale as $\hbar$. The error analogous
to Eq.~\ref{a2t} that we are interested in
for a compact phase space is the difference between performing a sum
over intermediate channels $q'$
and performing the corresponding {\it integral} by stationary phase.
We can then write
\begin{equation}
A_{2t}(i,k)=\sum_{j=0}^{N-1} A_t(i,j)A_t(j,k) + O(N^{-3/2}) \,.
\end{equation}
Note that the error scales as $N^{-3/2} \sim \hbar^{3/2}$ because normalization
(probability conservation) requires
that the actual matrix elements of $A_t$ and $A_{2t}$ scale as $N^{-1/2} \sim
\hbar^{1/2}$. The relative error is scaling as $N^{-1} \sim \hbar$.
We now define a natural norm for measuring the difference
between two matrices (under which a unitary matrix has norm one),
\begin{eqnarray}
\|A-B\|^2  & \equiv  & {1 \over N} {\rm tr} (A-B)^\dagger(A-B) \\
& = & {1 \over N} \sum_{ij} |(A-B)_{ij}|^2 \,. \nonumber
\end{eqnarray}
We then have 
\begin{equation}
\label{onemult}
\|A_{2t}-(A_t)^2\|^2 = O(\hbar^2) \,,
\end{equation}
whereas the matrices
$A_t$ individually have norm of order unity.
Of course, the coefficient in front of the $O(\hbar^2)$ depends on the
amount of nonlinearity in the underlying classical dynamics. For example,
the cat map given by
Eq.~\ref{catmap} is exactly linear, and produces a semiclassical
dynamics $A_t$ which is exactly multiplicative. This, however, is not very
interesting because in that case $A_t$ is also equal to the exact quantum
dynamics $U^t$. For generic perturbing potentials $V(q)$ or $T(p)$, the 
semiclassical answer differs form the quantum, and in that case the number
multiplying $\hbar^2$ will indeed be of order unity.

The result in Eq.~\ref{onemult} is already quite promising. It tells us that
for small $\hbar$, it is a very good approximation to compute the semiclassical
dynamics exactly for $10$ steps and then square the matrix instead of trying
to do the exact calculation for $20$ steps. The former is a much easier problem
to solve because the number of classical paths needed to compute $A_t$
scales exponentially with $t$. Inspired by this, we ask to what extent we
may approximate the exact time-$t$ semiclassical propagator for large $t$
by dividing $t$ into more and more shorter time intervals. Replacing $A_t$
by $(A_{t/M})^M$ involves making an error $M-1$ times, each time approximating
a stationary phase integration by exact multiplication. Assuming these errors
add incoherently, and taking $M$ to be large, we then have
\begin{equation}
\label{incoh}
\|A_t-(A_{t/M})^M\|^2 = O(M\hbar^2) \,.
\end{equation}
One can also obtain this result by iterating the procedure indicated by
Eq.~\ref{onemult} for $M$ that is a power of two. (In other words, we express
$A_{t}$ in terms of $A_{t/2}$ plus an error term, then $A_{t/2}$ in terms
of $A_{t/4}$ plus another error term, etc.)
However we need to be careful about the assumption of incoherent accumulation
of errors. Because the Hilbert space is finite-dimensional, eventually
we must consider interference between different error terms. To see this
in our formalism, let $B=A_{t/M}$ and $C=(A_t)^{1/M}$. (Although the $M$-th
root of a matrix is in general an ambiguous quantity, here there is no
ambiguity in what we mean by the matrix $C$:
we simply choose that root which
is closest to $B$.) Now the quantity that we are interested in is
$\|C^M-B^M\|^2$. Let $\epsilon\equiv C-B$. To lowest order in $\epsilon$,
\begin{eqnarray}
\label{manyeps}
\|C^M-B^M\|^2 & = & \|\epsilon B^{M-1} + B\epsilon B^{M-2} + \ldots +
B^{M-1} \epsilon  \nonumber \\  & + & O(\epsilon^2)\|^2 \,.
\end{eqnarray}
Now we work in the basis in which $B$ is diagonal, and write $\epsilon$ as
the sum of its diagonal and off-diagonal parts in that basis. The diagonal
part of $\epsilon$ commutes through $B$, giving a coherent contribution
from the $M$ terms in the sum. The off-diagonal part leads to an incoherent
contribution because the eigenphases of $B$ are generic.
So we obtain  for the basis-independent norm
\begin{eqnarray}
\|C^M-B^M\|^2  & = & O(M \|\epsilon_{\rm off-diag}\|^2)+
 O(M^2\|\epsilon_{\rm diag}\|^2) \nonumber \\
& + & O(M^2 \|\epsilon^2\|^2) + \ldots \,.
\end{eqnarray}
Now as long as the errors in the stationary phase approximation are not
{\it preferentially} diagonal (i.e. not in general tending to
multiply the exact answer), the weight
of the matrix $\epsilon$ that is on the diagonal is a fraction $1/N\sim\hbar$
of the total weight of $\epsilon$. Noting from Eq.~\ref{onemult} that 
$\|\epsilon\|^2=O(\hbar^2)$, we see that
$\|\epsilon_{\rm diag}\|^2=O(\hbar^3)$. Furthermore, for the values of $M$
that we are going to consider
(up to the Heisenberg time $T_H \sim \hbar^{-1}$),
terms higher order in $\epsilon$ (e.g. $O(M^2\hbar^4)$)
can be ignored. So we obtain
\begin{equation}
\label{multerror}
\|A_t-(A_{t/M})^M\|^2 = O(M\hbar^2) + O(M^2\hbar^3) \,,
\end{equation}
valid for $M\le O(\hbar^{-1})$.

\subsection{Heisenberg time dynamics}

In particular let us consider what happens at the Heisenberg time
$t =T_H \sim O(\hbar^{-1})$ (computing the semiclassical dynamics to times
longer than this will not produce interesting new information because 
individual eigenstates and eigenvalues will already have been resolved).
Let $T_Q$ be the ``quantization time", the time for which we will compute
the semiclassical propagator $A_{T_Q}$ exactly. 
First, letting $M=t/T_Q$, we can rewrite Eq.~\ref{multerror} as
\begin{equation}
\label{neweq}
\|A_t-A_{T_Q}^{t/T_Q}\|^2=O\left({t\hbar^2 \over T_Q}\right)
 +O\left({t^2\hbar^3 \over
T_Q^2}\right) \,.
\end{equation}
Then by the Heisenberg
time $t=O(\hbar^{-1})$, we accumulte an error
\begin{equation}
\label{twoterms}
\|A_{T_H}-(A_{T_Q})^{T_H/T_Q}\|^2=O\left({\hbar \over T_Q}\right)+
O\left({\hbar \over T_Q^2}\right) \,.
\end{equation}

The above formula is expected to hold for all
$T_Q \ge 1$, where classical paths
exist connecting any two coordinate points.
If we apply the result to the case $T_Q=1$, which characterizes the
one-step Bogomolny propagator, 
we find 
\begin{equation}
\label{scbogo}
\|A_{T_H}-(A_1)^{T_H}\|^2  =  O(\hbar) \,.
\end{equation}
So one result of the above calculation is that for smooth one-dimensional
kicked systems (or for two-dimensional Hamiltonian systems, where the scaling
of the density of states works identically), in the absence of caustics the
exact semiclassical propagator deviates from the iteration of the one-step
dynamics only by an error term of order $\sqrt\hbar$, by the Heisenberg
time. How does this propagation
relate to the exact quantum dynamics? For a single
time step we know that the relative error between semiclassical and quantum
amplitudes scales as $\hbar$, so
\begin{equation}
\|A_1-U\|^2=O(\hbar^2) \,.
\end{equation}
Following a line of reasoning completely analogous to the one that took us
from Eq.~\ref{onemult} to Eq.~\ref{scbogo} (i.e. noticing that off-diagonal
terms in the error add incoherently, etc.), we find
\begin{equation}
\|(A_1)^{T_H}-U^{T_H}\|^2 = O(\hbar) \,.
\end{equation}
Now combining this  with Eq.~\ref{scbogo} we obtain a relation between the
exact semiclassical and exact quantum answers:
\begin{equation}
\label{scquan}
\|A_{T_H}-U^{T_H}\|^2 = O(\hbar) \,.
\end{equation}
Thus, in the absence of discontinuities and caustics the semiclassical
approximation is expected to do
quite a good job in approximating the quantum dynamics
all the way out to the Heisenberg time (in two dimensions, that is -- 
in three dimensions the same analysis leads to the conclusion that the
semiclassical approximation is marginal at the Heisenberg time).

But there still is a difference between the long-time semiclassical and
quantum behavior, implying also a difference between the corresponding
eigenvalues and eigenstates. These corrections arise from ``hard quantum"
effects, those beyond the stationary phase approximation. To be able
to separate these from those effects which are purely semiclassical one
needs to be able to compute $A_t$ explicitly for $t\sim T_H$, to a better
approximation than that given by the exact quantum mechanics, Eq.~\ref{scquan}.
Iteration of the one-step semiclassical propagator $A_1$ will not do,
since it differs from the true semiclassical propagation as much as from the
quantum. So we proceed to consider a quantization time $T_Q \gg 1$. From
Eq.~\ref{twoterms} we see that iteration of $A_{T_Q}$ comes much closer to
the exact semiclassics than the difference between the latter and
quantum mechanics.
\begin{equation}
\|(A_{T_Q})^{t/T_Q}-A_t\|^2 \ll \|A_t-U^t\|^2 \,,
\end{equation}
for $t\sim T_H$. This implies
\begin{equation}
\|(A_{T_Q})^{t/T_Q}-A_t\|^2 \ll \|(A_{T_Q})^{t/T_Q}-U^t\|^2 \,,
\end{equation}
and thus iteration of $A_{T_Q}$ gives us an answer much closer to $A_t$
than to $U^t$, allowing us to see how the long-time semiclassical
dynamics differs from the quantum dynamics, without the need for
doing an exponentially
large amount of work. The approximation is a controlled one,
and one can keep increasing $T_Q$ until the desired level of convergence
to the exact semiclassics is reached.

These ideas, extended properly to quantization of non-compact phase
spaces, can be used for example to see dynamic semiclassical localization
without having to sum explicitly over a number of classical paths
that is exponential in the localization time. The results are presented
in a companion paper\cite{iterloc}, and address a long-standing
question in the literature over whether dynamic localization is a semiclassical
or hard quantum phenomenon.

\section{Numerical tests}
\label{numtest}

We proceed now to justify numerically the power-counting arguments
presented in the previous section. The system we will use is the
kicked system of Eq.~\ref{clasdyn}, with $m=n=1$, and $V(q)=
-(K /(2\pi)^2)\sin 2\pi q$. This can be thought of as a standard
map (kicked rotor) with an extra inverted harmonic oscillator potential
$-{1 \over 2}q^2$, or as
a sinusoidal perturbation of the $\left[\begin{array}{cc} 1 & 1 \\ 1 & 2
\end{array}\right]$ cat map. $T(p)$, the perturbation of
the quadratic kinetic term, is set to zero for simplicity. Also
for simplicity periodic boundary conditions, with no phases, are imposed, i.e.
$\epsilon_0=\epsilon_1=0$. The caustic-free
condition requires kick strength $|K|<1$. The system is then
guaranteed to satisfy the Anosov
property, as explained in the discussion following Eq.~\ref{jacob}. We
choose $N=256$ to be the dimension of the Hilbert space. This is well
in the semiclassical regime, and large
enough so that a direct summation of classical paths to the Heisenberg
time (there are $3^{256}$ of them) is clearly not practical. 

We then fix
a value of $K$, at $K=0.5$, and compute the matrix elements of the semiclassical
propagator between quantum states in the momentum basis (using $p$ is
natural because we are thinking of this as a kicked system; also
working in momentum space
leads naturally into the study of dynamical localization). Exact
semiclassical matrices $A_{T_Q}$ are computed for quantization times
$T_Q=1 \ldots 8$, using the Gutzwiller--Van-Vleck expression, Eq.~\ref{gvv}.
The convergence
of the iterative approximation can
then be investigated. We first compute the quantity
$\|(A_{T_Q})^{t/T_Q}-(A_8)^{t/8}\|^2$ for $T_Q=1,3,5,7$, and for a range of 
times $t$ that extends beyond the Heisenberg time $T_H=256$. 
As explained above, the exact semiclassical propagator for times of order 
$T_H$ cannot be computed exactly\footnote{At the end of the next section, we
will see an example of a numerical test for moderate time $t$, where exact
semiclassical calculation is in fact possible, though very time-consuming.
There, approximation techniques are explicitly shown to produce a very good
answer with much less effort.}
, so we use the difference between
approximations at different values of $T_Q$ as measure of convergence
as $T_Q \to \infty$. We assume errors in successive approximations are 
uncorrelated, i.e. 
\begin{equation}
\label{twoapprox}
\|A_{T_Q}^{t/T_Q}-A_{T'_Q}^{t/T'_Q}\|^2 \approx
\|A_{T_Q}^{t/T_Q}-A_t\|^2+\|A_{T'_Q}^{t/T'_Q}-A_t\|^2 \,.
\end{equation}
For comparison
purposes, the difference with the quantum dynamics $\|U_t-(A_8)^{t/8}\|^2$
is also computed. All these are plotted as a function of time $t$ in
Figure~\ref{fig1}. From top to bottom, the quantities plotted are the
squared differences between the eight-step iterated semiclassics
$A_8^{t/8}$ and
(i) the one-step iterated semiclassics $A_1^t$, (ii) the quantum mechanics
$U^t$, (iii) the three-step iterated semiclassics $A_3^{t/3}$, (iv)
the five-step iterated semiclassics, and finally (v) the seven-step iterated
semiclassics. Each set of points is also fitted to a function of the form
$at+bt^2$, as suggested by Eq.~\ref{neweq}.

We notice first of all that the three-step, five-step, and
seven-step approximation come progressively closer to the eight-step
approximation which is our basis of comparison. The differences between
all these are significantly smaller than that between any of these and the
quantum dynamics. Finally, the one-step iterated approximation does a
(relatively)
poor job of reproducing the long-time semiclassical behavior (however the
difference between it and the other calculations 
is still small, due to the smallness
of $\hbar$). Consistent with the prediction of
Eqs.~\ref{neweq},~\ref{twoapprox},
all the curves are well fitted by the sum of a linear and quadratic functions
of time. Furthermore, at least for $T_Q>3$, the linear term is seen to dominate
for times up to the Heisenberg time ($T_H=N=256$ in this calculation). Thus,
to understand the convergence of the iterative method for $T_Q \gg 1$, it is
sufficient to look at the behavior of the linear term in Fig.~\ref{fig1}.

This is in fact done in Fig.~\ref{fig2}. For each $T_Q$,
we find the numerical value of $a(T_Q)$ that fits
\begin{equation}
\|A_{T_Q}^{t/T_Q}-A_8^{t/8}\|^2=a(T_Q)t+b(T_Q)t^2
\end{equation}
In Fig.~\ref{fig2}, the quantity $a(T_Q)$ is plotted vs. $T_Q$, using plusses
for $N=256$ and crosses for $N=128$. However, for $T_Q=1$, the difference
between the {\it quantum} dynamics and the eight-step iterated semiclassics
is plotted, i.e the linear coefficient of $\|U^t-A_8^{t/8}\|^2$, {\it not}
$\|A_1^t-A_8^{t/8}\|^2$. The latter value is off the scale in the Figure, being
$30\cdot 10^{-7}$ for $N=256$ and $115\cdot 10^{-7}$ for $N=128$ (notice that
this error, though relatively large compared to those obtained for bigger $T_Q$,
still has the right $\hbar^2$ dependence).
From Eqs.~\ref{neweq},~\ref{twoapprox},
we predict
\begin{equation}
\label{atq}
a(T_Q)=c\hbar^2 ({1 \over 8} + {1 \over T_Q}) \,,
\end{equation}
for $T_Q \gg 1$, where $c$ is an undetermined constant of order unity. This
predicted behavior fits the observed values of $a(T_Q)$ quite nicely, with
$c=0.29$. The resulting curves, following Eq.~\ref{atq} are plotted in
Fig.~\ref{fig2}. We see that the error in the iterative approximation
indeed scales as $\hbar^2$  per time step, leading to an error of
order $\hbar$ by the Heisenberg time. Furthermore, the error goes to zero as
$T_Q \gg 1$, in the predicted manner, and is already much smaller than the
difference between the quantum mechanics and the semiclassics by $T_Q=3$.

\section{Effective one-step long-time propagator}

We have seen in the preceding sections how to obtain convergent
approximations to the semiclassical propagator for long times, including
times beyond the Heisenberg time. This dynamics can now be fourier
transformed to obtain local densities of states
for various initial
wavepackets, and from these the eigenstates and eigenvalues could be
extracted.
In particular, the semiclassical spectrum can be obtained by tracing
over the fourier transformed semiclassical evolution. The quantization
time $T_Q$ can be increased until the desired level of
convergence is reached, and the
eigenstates
and eigenvalues obtained in this way can finally be compared with those 
of the quantum matrix.

This, however, is a rather tedious process, requiring
the approximate semiclassical propagator to be evaluated at many values of
$t$, from short times to times well beyond $T_H$. Moreover, if we settle
on a fixed quantization time $T_Q$, and computed all long-time dynamics using
it, we will finally obtain nothing more than the eigenstates and eigenvalues
of the matrix $A_{T_Q}$. So clearly the sensible thing to do is to
diagonalize $A_{T_Q}$ directly for a range of values of $T_Q$ and check
for convergence as $T_Q \gg 1$. Unfortunately, a technical difficulty arises
here. Eigenstates of the true dynamics (quantum or semiclassical) with
eigenphases separated approximately 
by an integer multiple of $2\pi / T_Q$ may get mixed with each other in
the diagonalization of $A_{T_Q}$, preventing one from extracting the true
eigenstates. This behavior appears to be generic (it does not happen in $A_1$
due to level repulsion in the presence of chaos). One can try to get around
this difficulty by comparing the eigenstates of $A_{T_Q}$ for a range of $T_Q$,
and selecting those that come closest to agreeing for several values of $T_Q$.
Such a procedure does in fact appear to converge to a reasonable set of $N$
eigenstates, and it might be expected to produce results comparable to those
that would be obtained by fourier transforming long-time dynamics produced
by using several matrices $A_{T_Q}$\footnote{For example, we could approximate
$A_{100} \approx A_9^4A_8^8$, $A_{101} \approx A_9^5A_8^7$, etc. This procedure
is in fact useful for evaluating long-time propagators at arbitrary times
$t$, including those that are not divisible by a suitable value of $T_Q$.}.

However, a simpler solution now presents itself. Given that we believe
$A_{T_Q}$ have similar eigenstates for all values $T_Q \gg 1$, these must
also be the eigenstates of
\begin{equation}
\label{ddef}
D_{T_Q}\equiv A_{T_Q} A_{T_Q-1}^{-1}\,,
\end{equation}
for example.
No mixing of eigenstates should arise in $D_{T_Q}$ (because of level
repulsion), and moreover this method has the advantage that the semiclassical
eigenvalues can be read off directly, without having to decide which root
of an eigenphase must be taken in each case. $T_Q$ is then increased, and
convergence to the ``true" semiclassical eigenstates and eigenvalues
is obtained (convergence is expected based on what we know about the
convergence of the dynamics from the two previous sections).

In fact, it is sufficient to look at convergence of the $D_{T_Q}$ matrix.
In Fig.~\ref{fig3}, the results of such an analysis are presented,
for the same system as that studied numerically in Section~\ref{numtest}.
Here a Heisenberg time (matrix size) $N=64$ is used.
We plot the squared norm $\|D_{T_Q}-D_9\|^2$ for $2 \le T_Q \le 6$
(for $T_Q > 6$, numerical errors begin to play a role in the error
analysis, however, the downward trend continues to $5.4 \cdot 10^{-12}$
for the difference between $D_8$ and $D_9$). For
comparison, we also measure $\|A_1-D_9\|^2=4.6\cdot 10^{-5}$
and $\|U-D_9\|^2=6.2\cdot 10^{-6}$. 

Remarkably, we find exponentially fast convergence with the quantization
time, in marked contrast with the power law convergence obtained previously
for the dynamics. The data agrees well with the exponential form
$\hbar^2\exp(8.1-3.2T_Q)$, which is also plotted in Fig.~\ref{fig3}.
This exponential behavior is consistent with
analogous results seen in cycle expansion methods. How can it be reconciled
with the power-law behavior seen in Eq.~\ref{multerror}? Let 
\begin{equation}
\label{dstar}
D_\star \equiv \lim_{T_Q \to \infty} D_{T_Q} \,.
\end{equation}
Now clearly $A_t=D_\star^t(1+\epsilon_t)$, where $\epsilon_t$ is a
correction falling off exponentially with $t$, would satisfy the finding
(upon substitution into Eq.~\ref{ddef})
that $D_{T_Q}$ converges exponentially quickly. However, it would not be
consistent with Eq.~\ref{multerror}, which requires power-law behavior
for the dynamics, and which we have seen verified numerically.
The two results can be reconciled if
we notice that exponential convergence of $D_{T_Q}$ to $D_\star$ only requires
\begin{equation}
\label{estar}
A_t=D_\star^t(1+\epsilon_\star+\epsilon_t) \,.
\end{equation}
Here $\|\epsilon_t\|^2=O(\hbar^2 \exp{(-\alpha t)})$, while $\epsilon_\star$ is a
$t-$independent matrix with norm $\|\epsilon_\star\|^2=O(\hbar^2)$. 
$\alpha$ is a constant associated with the mixing time scale of the underlying
classical dynamics. Now
when taking the product $D_t\equiv A_t A_{t-1}^{-1}$, the $\epsilon_\star$
contribution cancels, producing $D_\star$ plus a
time-dependent correction that goes as $O(\hbar^2 \exp{(-\alpha t)})$. However,
if we compare $A_t$ with $A_{t/M}^M$ using Eq.~\ref{estar},
where $t$ and $t/M$ are assumed for simplicity to be large,
we notice that the latter quantity has an extra $M-1$ terms of order
$\epsilon_\star$ compared with the exact semiclassics $A_t$. This is
consistent with what we found in Eq.~\ref{manyeps}, and leads to power-law
errors as in Eq.~\ref{multerror}. For example, from Eq.~\ref{estar}
we see $A_{2t}-A_t^2=O(\epsilon_\star)$, in agreement with Eq.~\ref{onemult}.

Our interpretation of Eq.~\ref{estar} is that $\epsilon_t$ is an
error that results from
concatenating short classical paths to produce long ones. It is
thus analogous to the errors obtained in approximating long periodic orbits
by short ones in cycle expansion methods. $\epsilon_\star$, on the other
hand, is an error in some way associated with projecting the full
semiclassical dynamics for any time $t$ onto quantum initial and final
states. This error is therefore independent of the time after which
such projection is performed. It should be possible to make a connection
between this effect and what happens in the baker's map, for example,
when a slightly higher-dimensional effective semiclassical space (in which the
semiclassical dynamics is almost exactly iterative) must be projected onto $N-$vectors
to produce quantities that can be compared with quantum matrix
elements\cite{longbaker}.
One would also like to understand better the parametrization ambiguities
in the definition of $D_\star$ and $\epsilon_\star$. For example, we could have
written
\begin{equation}
A_t =(1+\tilde\epsilon_\star+\tilde\epsilon_t)\tilde D_\star^t
\end{equation}
as
an alternative to Eq.~\ref{estar}. This would be natural is we had looked
at the quantity $A_{T_Q-1}^{-1}A_{T_Q}$ ($=(A_{-T_Q}A_{1-T_Q}^{-1})^\dagger$)
instead of $D_{T_Q}$ as defined above.

In any case, we see that the matrix $D_\star$ is key to understanding the
long-time semiclassical dynamics. This matrix is an effective one-step
propagator that can be used to obtain the semiclassical propagator at
time $t+1$ with exponential accuracy, given the propagator for time $t$.
It thus gives to us in a trivial way the stationary properties of the
long-time semiclassical evolution.

To show how the factorization given in Eq.~\ref{estar} can be used in practice,
we take the kicked map studied earlier in this section and compute
an approximation to $A_9$ by evaluating
\begin{eqnarray}
\label{a9}
A_9 & \approx & D_\star^9(1+\epsilon_\star) \nonumber \\  & = & 
D_\star^4 D_\star^5(1+\epsilon_\star) \nonumber \\ & \approx  & D_\star^4 A_5
\nonumber \\ & \approx & (A_5 A_4^{-1})^4 A_5
\end{eqnarray}
All errors $\epsilon_t$ in Eq.~\ref{a9}
are exponentially small in the relevant times $t$:
an error of size $\epsilon_9$
is made in the first line, and errors of size $\epsilon_5, \epsilon_4$ in the
third and fourth lines, the latter of course
dominating the error in the final answer.
Notice that we need to evaluate the
semiclassical dynamics $A_t$ exactly at two values of $t$ to obtain the
two matrices $D_\star$ and $\epsilon_\star$. After computing the 
semiclassical propagator $A_9$ exactly, we find (factoring out the
$\hbar$ dependance), that the error $\|(A_5 A_4^{-1})^4 A_5-A_9\|^2$ is
$0.00003\hbar^2$. In the same units, we find the error $\|A_5 A_4 - A_9\|^2$
to be $0.5\hbar^2$, still not bad, but lacking the exponential error 
suppression.
For comparison, the difference between the exact semiclassical
$9-$step
propagator $A_9$ and the quantum mechanics $U_9$
is $3.1\hbar^2$, while the difference from
the $9-$times iterated one-step propagator ($\|A_1^9-A_9\|^2$) is
$89.5\hbar^2$. 
(It is interesting to note just how poorly iteration of the one-step
semiclassical propagator $A_1$ does in reproducing the correct
longer-time semiclassical behavior.)

Of course, this procedure can now be extended to
times much longer than $9$, where exact evaluation of the semiclassics
is clearly not practical. Six digits of accuracy in long-time semiclassical
propagation (beyond what is expected merely from the smallness of
$\hbar$) are obtained with only a modest amount of work.

\section{Conclusion}
\label{conclusion}

We have seen in the preceding section that long-time semiclassical
behavior is given to exponentially good accuracy by iteration of
an effective matrix $D_\star$
that is different from both the quantum evolution
matrix $U$
and the one-step semiclassical propagator $A_1$. In fact, the one-step
semiclassical propagator contains essentially no information about
long-time semiclassical dynamics, except of course for the similarity
based on both being related to the quantum evolution. However, knowing the
semiclassical dynamics for times $T_Q \gg 1$ (measured in units of the 
shortest periodic orbit time), enables one to deduce stationary semiclassical
behavior, including dynamical information to the Heisenberg time and beyond.
Why is long-time semiclassical information exhausted at $T_Q \gg 1$? In
periodic orbit methods, one expects knowledge of long-time dynamics to be
contained in orbits of length up to the mixing time ($\sim \log N$), at which
point phase space has been explored at the scale of Planck's constant. Longer
orbits can be produced from these if we allow smearing over sub-planck
coordinates. The $O(N)$ periodic points at the mixing time contain $O(N^2)$
pieces of information if we record where each lies in relation to some 
quantum basis.
Similarly, the full quantum theory can of course be described by an
$N \times N$ matrix. The information required to obtain the semiclassical
theory is only slightly bigger, scaling in the same way with $N$.
We note that collecting
and iterating semiclassical information after one time step on a scale
of $1/N$ allows one to resolve periodic orbits up to the mixing time, 
while collecting after a slightly longer exact evolution is equivalent
to having knowledge of periodic orbits longer than the mixing time, where
exponential convergence is expected.

The results obtained here may shed light on the accuracy of the Gutzwiller
trace formula in the energy domain. Though the relationship between the
two approaches is a nontrivial one (the trace formula focusing on {\it periodic}
orbits, and thus corresponding to
a stationary phase integration of the dynamics), one may hope that further
progress may be made in bringing together dynamical and periodic orbit
methods. In any case, we have already seen that the stationary phase
integration implicit in the trace formula is {\it not} necessary for obtaining
sensible semiclassical eigenvalues and eigenstates,
which compare well with those
of an exact quantum calculation.

More work is also needed in bringing together the approximation methods
discussed here with other semiclassical consolidation techniques, such
as those that have been used quite successfully for the baker's
map\cite{longbaker,iterbaker}.

These iterative methods may in addition
provide a new perspective on the long-standing questions
of classical periodic
orbit correlations and of how semiclassical dynamics can reproduce
Heisenberg-time quantum behavior\cite{pocorr}.
All such issues become much less mysterious
once we realize that long orbits are given to a good approximation by
concatenating shorter classical trajectories. Thus, long-time semiclassical
evolution can produce delta-function peaks in the spectrum for essentially the
same reason as the quantum mechanical evolution, i.e. it is given
(approximately) by iteration of an (almost) unitary finite matrix.

Many of the findings may be modified in the presence of discontinuities,
strong diffraction, and caustics, where classical-quantum correspondence
is more poorly understood. In particular, the proliferation of caustics 
may be expected to cause a divergence between semiclassical and quantum
dynamics well before the Heisenberg time is reached\cite{helleretc}. In these
situations, a better understanding of iterative dynamical methods may still
allow one to follow the semiclassical dynamics past the scale at which 
this divergence occurs, and to see explicitly when and in what way the
semiclassical approximation breaks down (and what corrections to the 
semiclassical formulas may be necessary to restore long-time
correspondence). In some cases, certain qualitative features of the
quantum dynamics (e.g. dynamical localization\cite{iterloc}, level spacing
statistics) may be well reproduced by the long-time semiclassical dynamics,
even when detailed correspondence between the propagators has been lost.

\section{Acknowledgements}

This research was supported by the National Science Foundation under
Grant No. 66-701-7557-2-30.
The author would like to thank the Isaac Newton Institute
for Mathematical Sciences, Cambridge, where some of the work was performed.
Much of the motivation for this project is the result of numerous discussions
with E. J. Heller, during which the concept of a quantization time was
developed.

\begin{figure}
\caption{
Convergence to long-time semiclassical behavior with increasing
quantization time $T_Q$. A long-time calculation with $T_Q=8$
is used as the reference. From top to bottom, the five sets of data
represent the squared difference between this calculation and
(i) the iterated one-step semiclassics, (ii) the quantum mechanics,
(iii) the three-step iterated semiclassics, (iv) the five-step iterated
semiclassics, and (v) the seven-step iterated semiclassics. The Heisenberg
time is $T_H=256$. Each curve is fit to the sum of a linear and quadratic
function of time, in accordance with Eq.~\ref{neweq}.
}
\label{fig1}
\end{figure}

\begin{figure}
\caption{
Convergence
of the linear coefficients in the previous Figure (error per time step),
with increasing quantization time $T_Q$. From left to right, the four
data points measure the divergence per unit time of the $T_Q=8$
calculation from (i) the quantum mechanics, (ii) the three-step
semiclassical approximation, (iii) the five-step approximation, and
(iv) the seven-step approximation. The upper and lower sets of points
correspond to Heisenberg times $N=128$ and $N=256$, respectively.
Theoretical curves corresponding to Eq.~\ref{atq}, with $c=0.29$,
are drawn through the data.
}
\label{fig2}
\end{figure}

\begin{figure}
\caption{
Exponential convergence of the quantity $D_{T_Q}=A_{T_Q}A_{T_Q-1}^{-1}$ towards
the {\it effective} one-step long-time semiclassical propagator
$D_\star$ (Eq.~\ref{dstar}). Data provided for $D_2 \dots D_6$; $D_9$
is used as the reference.
}
\label{fig3}
\end{figure}

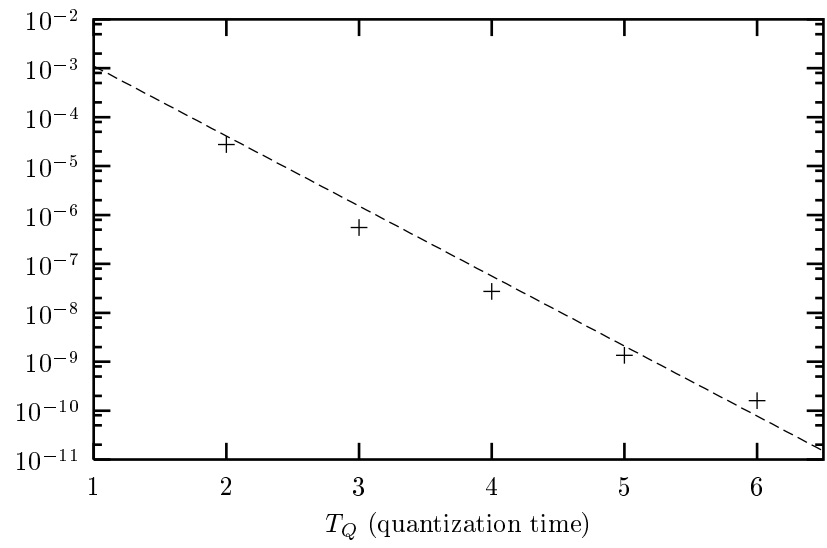
\end{document}